\documentclass[a4paper,12pt]{article}

\usepackage[utf8]{inputenc}
\usepackage{makeidx}  
\usepackage{graphicx}
\usepackage[utf8]{inputenc}
\usepackage{tikz}
\usetikzlibrary{shapes,arrows}
\usepackage[framemethod=TikZ]{mdframed}
\usepackage{amsfonts}
\usepackage{amsmath}
\newcommand{\Mod}[1]{\ (\text{mod}\ #1)}
\usepackage{rotating}
\usepackage{multirow}
\DeclareMathAlphabet{\mathpzc}{OT1}{pzc}{m}{it}
\usepackage{amssymb}
\usepackage{latexsym}
\usepackage{url}

\title{Compression-Based ECG Biometric Identification Using a Non-fiducial Approach}

\author{João M. Carvalho \and Susana Brás \and Armando J. Pinho}

\begin{document}
	\maketitle

	\begin{abstract}
		Due to its characteristics, there is a trend in biometrics to use the ECG signal
		for personal identification. Recent works based on compression models have shown that
		these approaches are suitable to ECG biometric identification. However, the best
		results are usually achieved by the methods that, at least, rely on one point of
		interest of the ECG.
		
		In this work, we propose a compression-based non-fiducial method, that uses a
		measure of similarity, called the Normalized Relative Compression -- a measure
		related to the Kolmogorov complexity of strings. Our method uses extended-alphabet
		finite-context models (\textrm{xaFCM}s) on the quantized first-order derivative
		of the signal, instead of using directly the original signal, as other methods do. 
		
		We were able to achieve state-of-the-art results on a database collected at the
		University of Aveiro, which was used on previous works, making it a good
		preliminary benchmark for the method.
		
	\end{abstract}

	\section{Introduction}
	The electrocardiogram (ECG) is a well-known and studied biomedical signal. To understand pathological characteristics, in clinical practice, it is usual to try to reduce the inter-variability that characterizes the signal. This inter-variability is precisely the source of richness that renders the ECG an interesting signal for biometric applications. Because of its desirable characteristics (universality, uniqueness, measurability, acceptability and circumvention avoidance \cite{Karimian2016}), it is suitable for biometric identification.

	We address this topic
	using a measure of similarity related to the Kolmogorov complexity, called the Normalized Relative Compression (\textrm{NRC}). To attain the goal, we use the generalized version of finite-context models (\textrm{FCM}), called extended-alphabet finite-context models (\textrm{xaFCM}) \cite{carvalho2017xafcm}, to represent each individual \cite{Bras2015a, carvalhoibpria2017, carvalho2017xafcm}---a compression-based approach that, besides ECG biometric identification, has been shown successful for different pattern recognition applications \cite{Pinho2011b, Pratas2014a,  Pinho2016a}. 
	
	In previous works, we have already used these methods \cite{Bras2015a, carvalhoibpria2017, carvalho2017xafcm}. However, the approach always relied on the detection of a fiducial point (a ``point of interest") in each heartbeat found in the ECG signal, called the \textrm{R-peak}. The detection of such points on clean signal is a computationally simple problem, with algorithms attaining accuracies of around 99.9\% \cite{Kathirvel2011}. But, as it is well known in biometrics, most times we need to deal with highly noisy signals, making that detection prone to error and, by transitivity, partially corrupting the whole process of identification.
	
	In this work, we present a non-fiducial method for ECG biometric identification, that uses a Lloyd-Max quantizer on first-order differentiation of the signal (the differences between consecutive points in the signal). We show that, using this approach, we improve previous state-of-the-art results obtained on a publicly available dataset\footnote{\url{http://sweet.ua.pt/ap/data/signals/Biometric_Emotion_Recognition.zip}}, originally collected for emotion classification \cite{Ferreira2016}. 
	
	The classification step uses 10 seconds of ECG signal before attempting classification. This choice was done based on the results achieved on a previous work \cite{carvalhoibpria2017}, where we showed that adding more time to the testing samples might not provide much of an advantage---and, of course, in a biometric system, we want the time needed before identification to be as small as possible.
	
	\subsection{Database}
	
	The database used in this work was collected \textit{in house}, where 25 participants were exposed to different external stimuli---\textit{disgust}, \textit{fear} and \textit{neutral}. Data were collected on three different days (once per week), at the University of Aveiro, using a different stimulus per day.
	
	The data signals were collected during 25 minutes on each day, giving a total of around 75 minutes of ECG signal per participant. Before being exposed to the stimuli, during the first 4 minutes of each data acquisition, the participants watched a movie with a beach sunset and an acoustic guitar soundtrack, and were instructed to try to relax as much as possible. 
	
	The ECG was sampled at 1000Hz, using the MP100 system and the software AcqKnowledge (Biopac Systems, Inc.). During the preparation phase, the adhesive disposable Ag/AgCL-electrodes were fixed in the right hand, as well as in the right and left foot. We are aware that such an intrusive set-up is not desirable for a real biometric identification system. However, since we have already used this database in previous works \cite{Bras2015a, carvalho2017xafcm, carvalhoibpria2017}, it provides a good benchmark for the methods against the previous approaches. 
	
	\subsection{Compression-based measures}
	
	Compression-based distances are tightly related to the Kolmogorov notion of complexity, also known as algorithmic entropy. Let $x$ denote a binary string of finite length. Its Kolmogorov complexity, $K(x)$, is the length of the shortest binary program $x^*$ that computes $x$ in a universal Turing machine and halts. Therefore, $K(x) = |x^*|$, the length of $x^*$, represents the minimum number of bits from which $x$ can be computationally retrieved \cite{KolmogorovA}. 
	
	The Information Distance (ID) and its normalized version, the Normalized Information Distance (NID), were proposed by Bennett \textit{et al.} almost two decades ago \cite{Bennett1998} and are defined in terms of the Kolmogorov complexity of the strings involved, as well as the complexity of one when the other is provided.

	However, since the Kolmogorov Complexity of a string is not computable, an approximation (upper bound) for it can be used by means of a compressor. Let $C(x)$ be the number of bits used by a compressor to represent the string $x$. We will use a measure based on the notion of \textit{relative compression} \cite{Pinho2016a}, denoted by $C(x||y)$, which represents the compression of $x$ relatively to $y$. 
	This measure obeys the following rules:
	
	\begin{itemize}
		\item $C(x||y) \approx 0$ iff string $x$ can be built efficiently from $y$;
		\item $C(x||y) \approx |x|$ iff $K(x|y) \approx K(x)$.
	\end{itemize}
	
	Based on these rules, the Normalized Relative Compression (\textrm{NRC}) of the binary string $x$ given the binary string $y$, is defined as
	\begin{equation}
	\textrm{NRC}(x||y)  = \frac{C(x||y)}{|x|},
	\end{equation}
	where $|x|$ denotes the length of $x$.
	
	A more general formula for the \textrm{NRC} of string $x$, given string $y$, where the strings $x$ and $y$ are sequences from an alphabet $\mathcal{A} = \{ s_1, s_2, \dots s_{|\mathcal{A}|}\}$, is given by
	\begin{equation}
	\textrm{NRC}(x||y)  = \frac{C(x||y)}{|x| \log_2{|\mathcal{A}|}}.
	\end{equation}
	
	\subsection{Extended-Alphabet Finite-Context Models}
	
	Let $\mathcal{A} = \{ s_1, s_2, \dots s_{|\mathcal{A}|}\}$ be the alphabet that describes the objects of interest.
	An extended-alphabet finite-context model (\textrm{xaFCM}) complies to the Markov property, i.e., it estimates the probability of the next sequence of $d > 0$ symbols of the information source (depth-$d$) using the $k > 0$ immediate past symbols (order-$k$ context).
	Therefore, assuming that the $k$ past outcomes are given by $x_{n-k+1}^{n} = x_{n-k+1} \cdots x_{n}$, the probability estimates, $P(x_{n+1}^{n+d}|x_{n-k+1}^{n})$ are calculated using sequence counts that are accumulated, while the information source is processed,
	
	\begin{equation}\label{eq_fcm2}
	P(w|x_{n-k+1}^{n}) = \frac{ v(w|x_{n-k+1}^{n}) + \alpha }{  v(x_{n-k+1}^{n}) + \alpha|\mathcal{A}|^{d} },
	\end{equation}
	where $\mathcal{A}^d = \{ w_1, w_2, \dots w_{|\mathcal{A}|},\ldots w_{|\mathcal{A}|^d}\}$ is an extension of alphabet $\mathcal{A}$ to $d$ dimensions, 
	$v(w|x_{n-k+1}^{n})$ represents the number of times that, in the past, sequence $w \in {\mathcal{A}^d}$
	was found having $x_{n-k+1}^{n}$ as the conditioning context and where
	\begin{equation}
	v(x_{n-k+1}^{n}) = \sum_{a \in \mathcal{A}^{d}} v(a|x_{n-k+1}^{n}) 
	\end{equation}
	denotes the total number of events that has occurred within context $x_{n-k+1}^{n}$. 
	
	In order to avoid problems with ``shifting" of the data, the sequence counts are performed symbol by symbol, when learning a model from a string.
	
	Parameter $\alpha$ allows controlling the transition from an estimator initially assuming a uniform distribution to a one progressively closer to the relative frequency estimator. In this paper we will use the parameter $\alpha$ chosen on ``auto" (for more details, please check the original paper \cite{carvalho2017xafcm}).
	
	The theoretical information content average provided by the $i$-th sequence of $d$ symbols from the original sequence $x$, is given by
	
	\begin{equation}
	- \log_2 P(X_i = t_i  | x_{id-k}^{id-1}) \text{ bits,} 
	\end{equation}
	where $t_i = x_{id}, x_{id+1} \cdots x_{(i+1)d-1}$.
	
	After processing the first $n$ symbols of $x$, the total number of bits generated by an order-$k$ with depth-$d$ \textrm{xaFCM} is equal to
	
	\begin{equation}
	- \sum_{i=1}^{n/d} \log_2 P(t_i|x_{di-k}^{di-1}),
	\end{equation}
	where, for simplicity, we assume that $n \Mod d = 0$.
	
	For compressing the first $k$ symbols of a sequence, because we do not have enough symbols to represent a context of length $k$, we always assume that the sequence is ``circular". For long sequences, specially using small contexts/depths, this should not make much difference in terms of compression, but as the contexts/depths increase, this might not be always the case.
	
	Since the purpose for which we use these models is to provide an approximation for the number of bits that would be produced by a compressor based on them, whenever we use the word ``compression", in fact we are not performing the compression itself. For that, we would need to use an encoder, which would take more time to compute. It would also be needed to add some side information for the compressor to deal with the circular sequences---but that goes out of scope for our goal.

	\subsection{Lloyd-Max Quantization}	
	Quantization is widely used in signal processing. It is a process that takes a signal and produces only a, usually predefined, discrete set of values. It is a very simple process. However, the design of the quantizer has a significant impact on the amount of compression obtained and loss incurred in a lossy compression scheme \cite{introdatacompression}.
	
	As mentioned before, we do not want to achieve real compression---we just want to compute measures of dissimilarity. For that reason, we can afford to use a lossy-compression scheme. We have loss of information by performing quantization on the ECG signal.
	
	We will refer to number of discrete intervals in the quantization as the \textit{alphabet size}. There is a fundamental trade-off to take into account while performing the choice of the alphabet size: the quality produced versus the amount of data necessary to represent the sequence \cite{Reading2013}. In this work, we have always used an alphabet size of 17, which the quantizer represents by the symbols corresponding to the first 17 letters of the alphabet: `A', `B', ... `P', `Q'.
	
	When the random variable to be quantized does not follow a uniform distribution, a nonuniform quantization should be performed. In order to decrease the average error of the quantization, we can try to approximate the input better in regions of high probability, perhaps at the cost of worse approximations in regions of lower probability. We can do this by making the quantization intervals smaller in those regions that have more probability mass \cite{introdatacompression}.	
	
	One example of such a quantization is a Lloyd-Max quantization, which can be useful when the distribution of the variable to quantize is given by some complex mathematical function, like the ECG signal, for which we can not find a simple mathematical function that describes the signal.

	\section{Method}
	
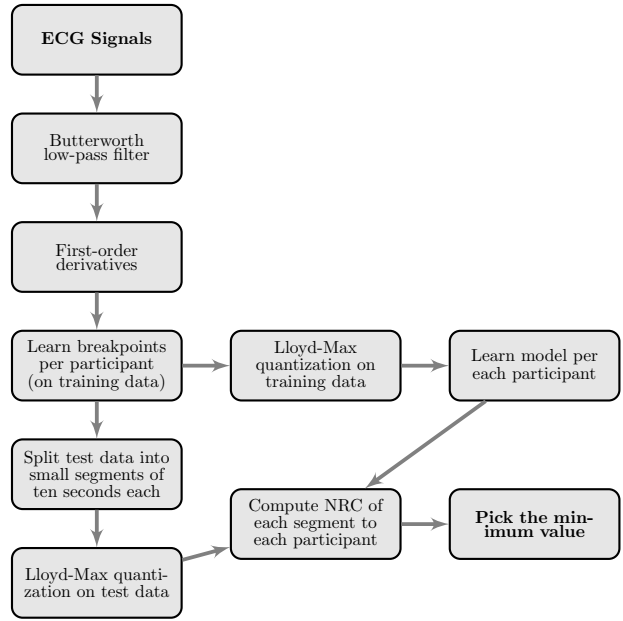
\begin{figure}
		\vspace*{0.5\baselineskip}
		
		\centering
		\tikzstyle{block} = [rectangle, draw, fill=black!10,
		text width=9em, text centered, rounded corners, minimum height=4em]
		\tikzstyle{line} = [draw, line width = 0.5mm, color=black!50, -latex']
		\tikzstyle{line_tick} = [draw, line width = 0.1mm, color=black!50, -latex']
		\begin{center}
			\begin{tikzpicture}[thick,scale=0.6, every node/.style={transform shape}, node distance = 2.4cm]
			\fontsize{11pt}{0}
			
			\node [block] (block00) {\textbf{ECG Signals}};
			
			\node [block, below of=block00] (block0) {Butterworth low-pass filter};
			
			\node [block, below of=block0] (block1) {First-order derivatives};
			
			\node [block, below of=block1] (block2) {Learn breakpoints per participant (on training data)};
			
			\node [block, below of=block2] (block4) {Split test data into small segments of ten seconds each};
			
			\node [block, below of=block4] (block5) {Lloyd-Max quantization on test data};
			
			\node [block, right of=block2, node distance=4.8cm] (block9) {Lloyd-Max quantization on training data};
			
			\node [block, right of=block9, node distance=4.8cm] (block6){Learn model per each participant};
			
			\node [block, below of=block9, node distance=3.5cm] (block7) {Compute NRC of each segment to each participant};
			
			\node [block, right of=block7, node distance=4.8cm] (block8) {\textbf{Pick the minimum value}};

			\path [line] (block00) -- (block0);
			\path [line] (block0) -- (block1);
			\path [line] (block1) -- (block2);
			\path [line] (block2) -- (block4);
			\path [line] (block2) -- (block9);
			\path [line] (block4) -- (block5);
			\path [line] (block6) -- (block7);
			\path [line] (block7) -- (block8);	
			\path [line] (block5) -- (block7);
			\path [line] (block9) -- (block6);
			\end{tikzpicture}
		\end{center}
		\caption{Overview of the method used in this work.} \label{fig_workflow1}
	\end{figure}

	An overview of the method used in this work can be seen in Fig.~\ref{fig_workflow1}. We start by cleaning the ECG signals by using a Butterworth low-pass filter of order 5 with a cutoff frequency of 30Hz. The obtained signal is then transformed into a series of differences (which corresponds to the first-order derivative of the signal).
	
	Since we want to apply a Lloyd-Max quantizer to this series, we perform a 2-pass process on the training data: first, for each participant in the database, we learn the breakpoints that optimize its Lloyd-Max
	quantization\footnote{The source code for the Python quantizer is publicly available under the GPL v3 license at \url{https://github.com/joaomrcarvalho/diffquantizer}};
	on the second phase, we apply the corresponding breakpoints to each participants' training data, in order to perform the quantization.
	
	From the quantized training data, it is possible to learn a model that describes each participant's data by using a context-based compressor, such as a \textrm{xaFCM}\footnote{The source code for the Python implementation of the extended-alphabet finite-context model based compressor is publicly available under the GPL v3 license at \url{https://github.com/joaomrcarvalho/xafcm}}. It is important to notice that each model, besides learning the \textrm{xaFCM}s, also takes notice on which participant it is representing---this is important, because those breakpoints will also be used during the testing phase.
	
	The splitting of the test data into segments of 10 seconds is then performed. At this point, it is only required to compute the amount of bits it takes to compress each of those segments, by each of the participants' models. This step is done in two phases: the first is to perform the quantization of the segment being tested using the breakpoints corresponding to the model that we are using; afterwards, the estimation of the amount of bits needed to represent that sequence using the \textrm{xaFCM} is computed. The model that produces less bits, i.e., the one which has a lower \textrm{NRC}, is our guess as the correct participant.

	\section{Experimental Results}
	
	We tried to replicate as much as possible the experimental setup used in previous works \cite{carvalho2017xafcm, carvalhoibpria2017}, in order to have fair benchmarks against those systems. However, since the previous methods used R-peak detection, the way to measure the size of the ECG samples used for testing was done in complete heartbeat cycles, instead of seconds. In those previous methods, we have used 10 heartbeats for each test. In order for the results to be comparable, in this work we assumed that one heartbeat is approximately 1 second. Therefore, each test is performed using 10 seconds of ECG data. Even if this approximation is not completely accurate, i.e., if we use a little more data (or less, depending on the heartbeat rate of each participant) than the previous experiments, it should not impact the results significantly, as we showed in \cite{carvalhoibpria2017}.
	
	All the experiments were performed on a Amazon AWS EC2 instance (c5.9xlarge), with a 3.0 GHz Intel Xeon Platinum (34 cores) CPU and 72GB of RAM. The operating system used was Ubuntu Server 16.04, and Python 3.6.4. The process could run on a regular laptop computer with 8GB of RAM, but we decided to use a cloud instance computer in order to make use of the parallelized code and have faster results. As mentioned in the previous section, all the base source code is freely available and can be downloaded from the Github repository.
	
	Using higher values of the depth, $d$, of the \textrm{xaFCM}, has the advantage of providing very fast results, at sometimes the possible cost of some decrease in accuracy (the theoretical explanation  for these concepts can be found in \cite{carvalho2017xafcm}). For that reason, we use high values of $d$ for finding the areas of interest for the parameter $k$ and, afterwards, we start decreasing the value $d$ and reduce the number of simulations that we need to run in order to find the optimal values of the context, $k$, in order to obtain more accurate results. Of course, the optimal values of $k$ also depend on the depth, $d$, but using high values of $d$ gives an idea of the area experiments should be performed more extensively.
	
\begin{figure}[!htb]
		\centering
		\vspace*{0.4\baselineskip}
		
		\begin{minipage}[b]{0.8\textwidth}
			\includegraphics[width=\textwidth]{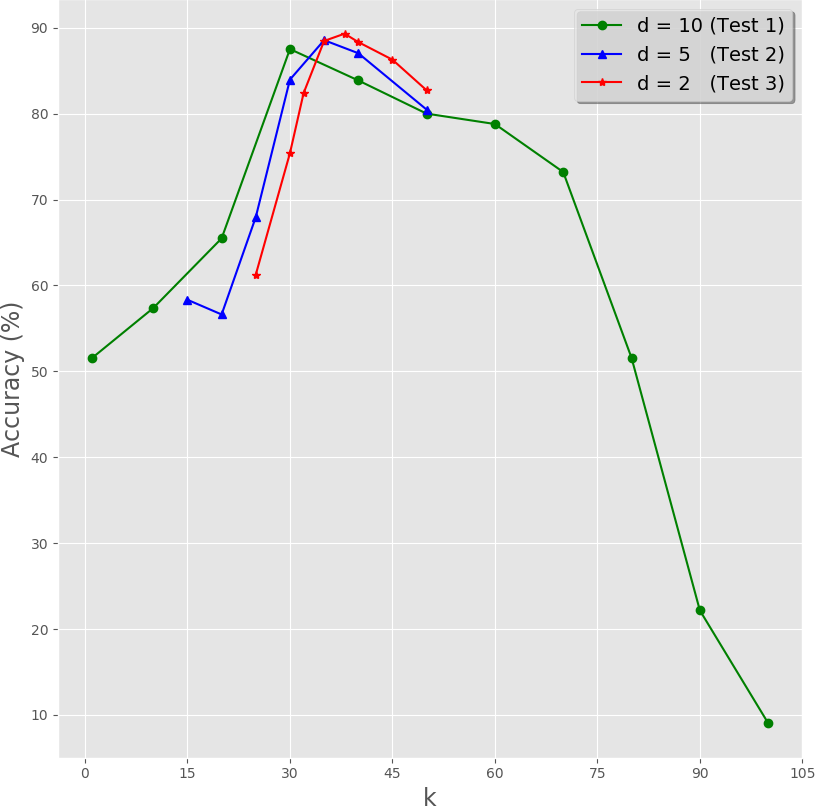}
			\caption{Biometric identification accuracy, using $d=2$, as a function of the parameter $k$.}\label{fig_biometry_k_changing_psic}
		\end{minipage}
	\end{figure}

	Figure~\ref{fig_biometry_k_changing_psic} shows all the experiments ran for biometric identification on this database, for different values for the depth $d$ and context $k$. As mentioned in the Method section, all the experiments used two days for training the models and all the available ten-second samples of ECG from the remaining day as the tests.
	
	The first phase of tests was ran with $d = 10$, experimenting contexts $k$ from $1$ up to $100$. It is easy to see from the plot marked in green (Fig.~\ref{fig_biometry_k_changing_psic}), that the possible area of interest for $k$ lays somewhere between $15$ and $50$---the best value was found for $k = 30$, with an accuracy of $87.5\%$. The second phase (marked as blue) uses $d = 5$ and narrows down the area of interest for $k$ from around 25 to 50, with the best results for $k = 35$, with an accuracy of $88.6\%$. Then, since we have a small region of interest, we performed some tests using $d = 2$ and the best value found was with $k = 38$, with an accuracy of $89.3\%$. Actually, if we look at the differences in accuracy, depending on the requirements in terms of speed, it might not even be worth using small values of $d$ for this application. The results in terms of the choice of $d$ are consistent with our previous results \cite{carvalho2017xafcm}. Regarding the context, $k$, from these results, we can infer that this new approach requires higher values of $k$ in order for the model to have a good internal representation of each participant. This might have an impact in terms of memory usage and time of execution. However, since the size of the data used is usually not significant (a couple of megabytes), the models also do not grow exponentially to values that can not be represented by a regular laptop, as they would with data like DNA sequences \cite{carvalho2017xafcm}.
	
\begin{figure}[!htb]
		\centering
		
		\vspace*{0.5\baselineskip}
		
		\begin{minipage}[b]{0.8\textwidth}
			\includegraphics[width=\textwidth]{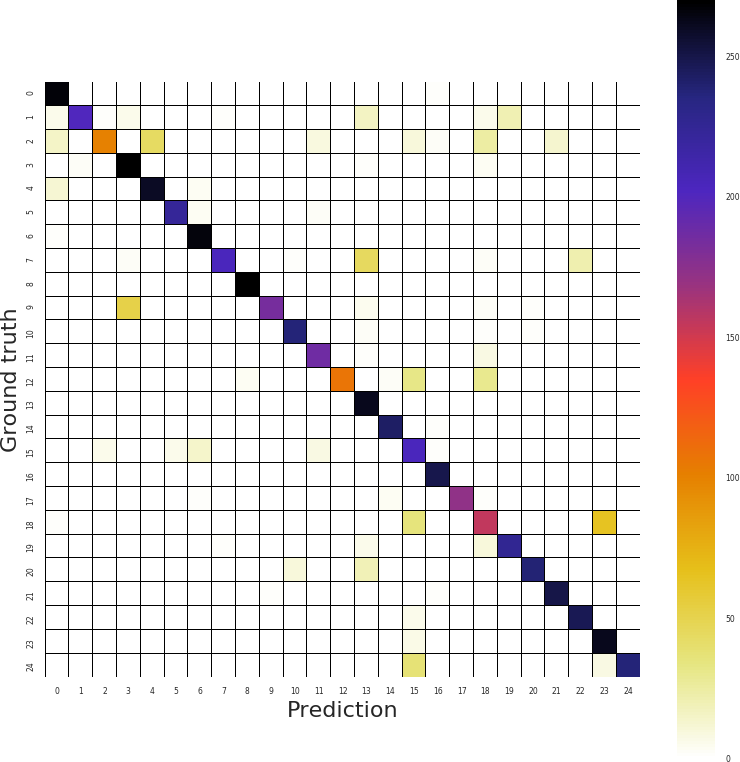}
			\caption{Confusion matrix for biometric identification using a \textrm{xaFCM} of context $k = 35$ and depth $d = 2$. This test used two days for training and the other day for testing. Each test was performed using 10 seconds of ECG. This experiment achieved an accuracy of $88.5\%$ and F1-score of $0.88$.
			}\label{fig_confusion_matrix_psic}
		\end{minipage}
	\end{figure}

	Besides the accuracy, it also useful to check a measure that takes into account the precision obtained. For that reason, and also to have a more clear understanding of the types of error that our system is performing for this dataset (to answer questions like ``how many false positives/true negatives are we obtaining in each class?"), we show the confusion matrix of the predictions made by the system, against the true labels. In Fig.~\ref{fig_confusion_matrix_psic}, we show the confusion matrix obtained for the experiment using $k=35$ and $d=2$, for which we obtained an accuracy of $88.5\%$ and F1-score of $0.88$.
	
	It is interesting to notice that when testing ECG signal belonging to participants 0, 3, 6, 8, 13, 14, 16 and 21, the system almost does not make any mistake. For the other participants, the system makes mistakes, but they are ``spread'' amongst different other participants, i.e., the system never consistently mistakes one participant by a specific different one. This is a very important feature on a biometric system, because it makes it harder for someone to fake a specific identity. On the first experiments performed with previous approaches, this was a major problem that we had. Our current justification is that this usually happens when the amount of training data provided for one participant is not proportional to the other ones---however, more research needs to be done to verify this claim.
	
	\section{Conclusions and Future Work}
	
	We have introduced a compression-based method, that works with first order derivatives, instead of the original signal, for performing ECG biometric identification.
	
	This method beats previous state-of-the-art methods using this database, achieving an accuracy of 89.3\%. Moreover, it uses the same amount of training data as the previous methods, that have attained, at most, around 80\% of accuracy \cite{carvalhoibpria2017}.
	
	
	We are confident that these results can be further improved. However, since the purpose was to introduce the main ideas associated to the method, we did not perform an exhaustive search for optimal parameters, neither experimented with mixtures of finite-context models (collaborative models), which, on machine learning terms, behave like a dynamic voting system.
	
	While these results seem very promising, future work needs to be done in order to check how well this approach works when dealing with intruders in the system, for example. For attaining that goal, the first step is to switch from a classification problem to a real biometric system, where there should be a threshold value for the \textrm{NRC}, instead of always accepting the minimum value as the correct participant. We intend to perform this change in a near future and also benchmark the method against other state-of-the-art ECG biometric identification methods, regarding the most significant databases available (namely, UoftDB \cite{uoftdb_original} and CYBHi \cite{DaSilva2014}).

\bibliographystyle{acm}
\bibliography{library}

\begin{thebibliography}{10}

\bibitem{Bennett1998}
{\sc Bennett, C., G{\'{a}}cs, P., and Li, M.}
\newblock {Information distance}.
\newblock {\em IEEE Transactions on Information Theory 44}, 4 (1998), 1407 --
  1423.

\bibitem{Bras2015a}
{\sc Br{\'{a}}s, S., and Pinho, A.~J.}
\newblock {ECG biometric identification: A compression based approach}.
\newblock In {\em 2015 37th Annual International Conference of the IEEE
  Engineering in Medicine and Biology Society (EMBC)\/} (aug 2015),
  pp.~5838--5841.

\bibitem{carvalhoibpria2017}
{\sc Carvalho, J.~M., Br{\~{a}}s, S., Ferreira, J., Soares, S.~C., and Pinho,
  A.~J.}
\newblock {Impact of the Acquisition Time on ECG Compression-Based Biometric
  Identification Systems}.
\newblock In {\em Pattern Recognition and Image Analysis. IbPRIA 2017. Lecture
  Notes in Computer Science, vol 10255. Springer, Cham\/} (jun 2017), Springer,
  Cham, pp.~169--176.

\bibitem{carvalho2017xafcm}
{\sc Carvalho, J.~M., Br{\'{a}}s, S., Pratas, D., Soares, S.~C., Ferreira, J.,
  and Pinho, A.~J.}
\newblock {Extended-Alphabet Finite-Context Models}.
\newblock {\em (Submitted)\/} (2017).

\bibitem{DaSilva2014}
{\sc da~Silva, H.~P., Louren{\c{c}}o, A., Fred, A., Raposo, N., and Aires-de
  Sousa, M.}
\newblock {Check Your Biosignals Here: A new dataset for off-the-person ECG
  biometrics}.
\newblock {\em Computer Methods and Programs in Biomedicine 113}, 2 (2014),
  503--514.

\bibitem{Ferreira2016}
{\sc Ferreira, J., Br{\'{a}}s, S., Silva, C.~F., and Soares, S.~C.}
\newblock {An automatic classifier of emotions built from entropy of noise}.
\newblock {\em Psychophysiology\/} (2016).

\bibitem{Reading2013}
{\sc {Gonzalez C.}, R., and Woods, R.~E.}
\newblock {Sampling and Quantization}.
\newblock In {\em Digital Image Processing}. Prentice Hall PTR, 2007.

\bibitem{Karimian2016}
{\sc Karimian, N., Wortman, P.~A., and Tehranipoor, F.}
\newblock {Evolving authentication design considerations for the internet of
  biometric things (IoBT)}.
\newblock In {\em Proc. Elev. IEEE/ACM/IFIP Int. Conf. Hardware/Software
  Codesign Syst. Synth. - CODES '16\/} (New York, New York, USA, 2016), ACM
  Press, pp.~1--10.

\bibitem{Kathirvel2011}
{\sc Kathirvel, P., Sabarimalai, M., Prasanna, S. R.~M., and Soman, K.~P.}
\newblock {An Efficient R-peak Detection Based on New Nonlinear Transformation
  and First-Order Gaussian Differentiator}.
\newblock {\em Cardiovascular Engineering and Technology 2}, 4 (oct 2011),
  408--425.

\bibitem{KolmogorovA}
{\sc Li, M., and Vit{\'{a}}nyi, P.}
\newblock {\em {An introduction to Kolmogorov complexity and its
  applications}}, 3rd~ed.
\newblock Springer, 1997.

\bibitem{Pinho2011b}
{\sc Pinho, A., and Ferreira, P.}
\newblock {Image similarity using the normalized compression distance based on
  finite context models}.
\newblock {\em 18th IEEE International Conference on Image Processing\/}
  (2011).

\bibitem{Pinho2016a}
{\sc Pinho, A.~J., Pratas, D., and Ferreira, P. J. S.~G.}
\newblock {Authorship Attribution using Compression Distances}.
\newblock In {\em Data Compression Conference\/} (2016).

\bibitem{Pratas2014a}
{\sc Pratas, D., and Pinho, A.~J.}
\newblock {A Conditional Compression Distance that Unveils Insights of the
  Genomic Evolution}.
\newblock In {\em 2014 Data Compression Conference\/} (mar 2014), IEEE,
  pp.~421--421.

\bibitem{introdatacompression}
{\sc Sayood, K.}
\newblock {\em {Introduction to data compression}}.
\newblock Morgan Kaufmann Publishers, 2000.

\bibitem{uoftdb_original}
{\sc Wahabi, S., Pouryayevali, S., Hari, S., and Hatzinakos, D.}
\newblock {On Evaluating ECG Biometric Systems: Session-Dependence and Body
  Posture}.
\newblock {\em IEEE Transactions on Information Forensics and Security 9}, 11
  (nov 2014), 2002--2013.

\end{thebibliography}

\end{document}